\documentstyle[prl,epsfig,aps,amsmath,amssymb,multicol]{revtex}

\begin{document}


\title{High performance of potassium n-doped carbon nanotube field-effect transistors}

\author{M.~Radosavljevi\'c$^{\rm a}$, J.~Appenzeller$^{\rm b}$, and Ph.~Avouris$^{\rm c}$}

\address{IBM Research Division, T.~J.~Watson Research Center, Yorktown Heights, New York 10598, USA}

\author{J. Knoch}

\address{Institut f\"ur Schichten und Grenzfl\"achen, Forschungszentrum J\"ulich, D-52425 J\"ulich, Germany}

\date{\today}

\maketitle

\begin{abstract}
We describe a robust technique for the fabrication of high
performance vertically scaled n-doped field-effect transistors
from large band gap carbon nanotubes. These devices have a tunable
threshold voltage in the technologically relevant range
($-1.3\,$V$\le V_{\rm th} \le 0.5\,$V) and can carry up to
$5$-$6\,\mu$A of current in the on-state. We achieve such
performance by exposure to potassium (K) vapor and device
annealing in high vacuum. The treatment has a two-fold effect to:
(i) controllably shift $V_{\rm th}$ toward negative gate biases
via bulk doping of the nanotube (up to about $0.6\, e^-/$nm), and
(ii) increase the on-current by $1$-$2$ orders of magnitude. This
current enhancement is achieved by lowering external device
resistance due to more intimate contact between K metal and doped
nanotube channel in addition to potential reduction of the
Schottky barrier height at the contact.
\end{abstract}

\vspace*{-.8cm}

\begin{multicols}{2}

\narrowtext

%
%

Recent fabrication of complimentary logic gates based on carbon
nanotube transistors (CNFETs) has intensified the interest in
nanoelectronic applications for these
materials.\cite{vd_nl,zhou_apl} These initial gates showed
performance which was limited by the device characteristics of the
constituent CNFETs. In the interim, many more improvements have
been made on p-CNFETs than on complimentary n-type
transistors.\cite{sw_apl,dai_nmat,dai_nature} This is mostly
because p-CNFETs are readily available under ambient conditions
due to oxygen induced dipoles at the interface between the metal
electrode and the CN.\cite{nl_dong} In contrast, n-CNFETs require
controlled environment, either by doping with electropositive
elements, such as potassium (K) metal \cite{bock_prb,dai_apl} or
by annealing/out gassing the oxygen at the
contacts.\cite{dai_nmat,vd_apl} While these two methods use very
different physical mechanisms to achieve n-CNFETs, neither has
been successful at approaching the performance (i.e.~on- and
off-states as well as subthreshold slope) of best p-CNFETs.

In this letter we report on a technique that allows reproducible
optimization of device characteristics by (i) increasing the
charge density in the channel and (ii) lowering of the external
series resistance. After outlining the experimental details of the
technique which consists of a combination of K-doping and device
annealing, we discuss the performance gains in both the on- and
the off-states. Our n-CNFETs show on-currents ($\sim 5$-$6\,\mu$A,
Ref.~\cite{vd_foot}) which rival those of the best p-CNFETs
produced to date\cite{sw_apl} from small diameter CNs as discussed
below.\cite{nanotubes} In the off-state, we obtain a value of
subthreshold swing $(S) \le 150\,$mV/decade due to good
electrostatic control\cite{ja_SBprl} by the gate field (oxide
thickness, $t_{\rm ox} = 5\,$nm). Both of these observations are
in agreement with expectations based on the symmetric band
structure and identical electron and hole masses in
CNs,\cite{avouris_book} and they remain unchanged as the device
threshold voltage is tuned over a technologically relevant gate
range, $-1.3\,$V$\le V_{\rm th} \le 0.5\,$V (corresponding to a
charge density increase from nearly zero up to about $0.6\,
e^-/$nm.)

\begin{figure}[ht]
 \begin{center}
 \begin{minipage}{8cm}
   \epsfclipon
   \epsfxsize=7cm
   \begin{center}
   \epsffile{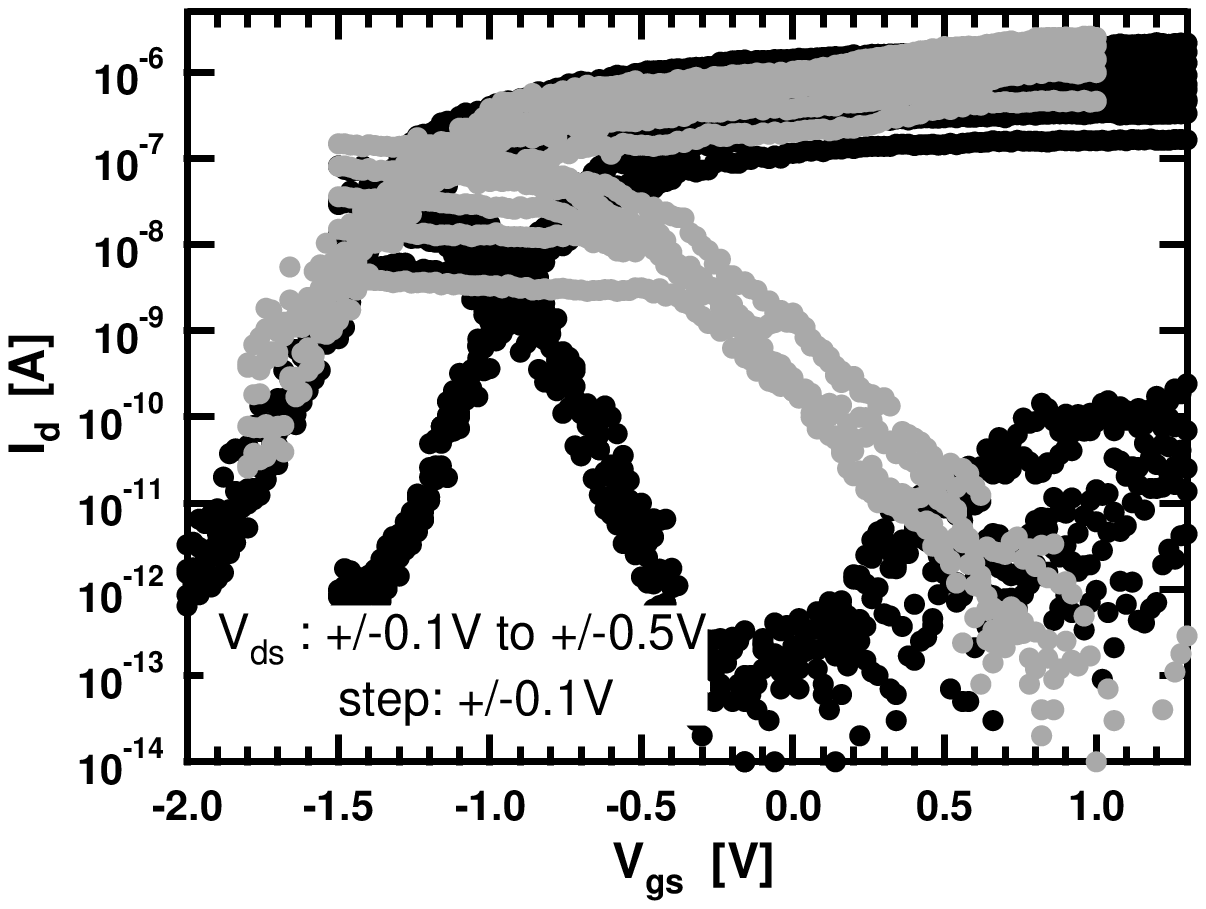}\\
   \epsfxsize=7cm
   \epsffile{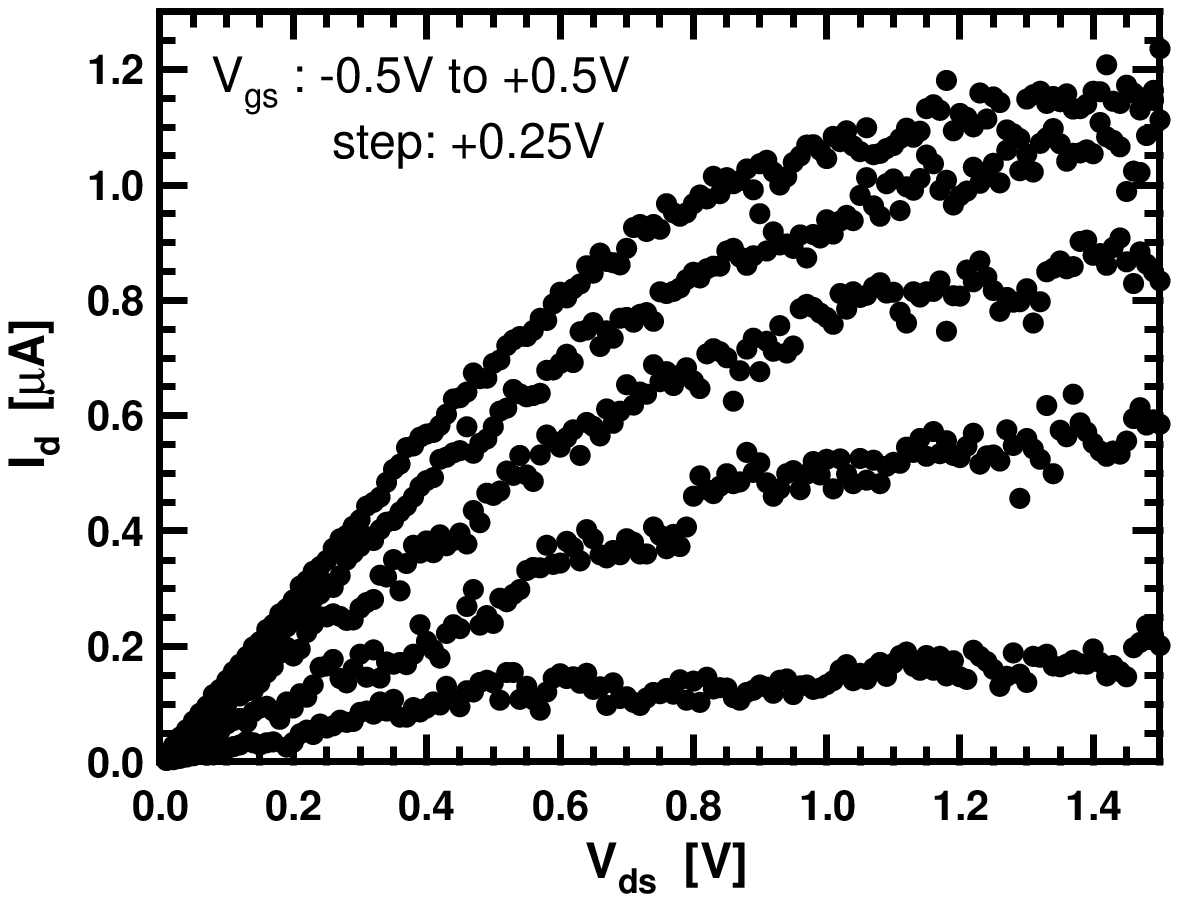}
   \end{center}
   \caption{(a) Transfer characteristics showing conversion
   of two different CNFETs (in black and grey) from p- to
   high performance n-type devices through exposure to K
   vapor and vacuum annealing. (b) Output characteristics at an
   intermediate stage of conversion (black) with transconductance
   in excess of $1\,\mu$S at $V_{\rm ds} = +1.5\,$V.}
   \label{Fig1}
 \end{minipage}
 \end{center}
\end{figure}

Finally, we discuss the connection of the observed changes in the
device characteristics to the cycle of K exposure/annealing,
focusing in particular on the effect of doping of the CN channel,
and lowering of the external device resistance due to formation of
a more intimate contact between potassium and the nanotube.

For this work, we have fabricated CNFETs with titanium source and
drain electrodes separated by 300-$400\,$nm and Si-backgate
($t_{\rm ox} = 5\,$nm) as described elsewhere.\cite{sw_jvst}
Specifically, we focus on rather small diameter nanotubes
($1.4\pm0.2\,$nm) which exhibit band gaps of around
$0.6$-$0.7\,$eV.\cite{nanotubes} These semiconducting CNs are
sufficiently similar to silicon in terms of their band gap size
that they are suitable for ultimately scaled transistor
applications. All transport data are acquired at negative
(positive) drain bias ($V_{\rm ds}$) for p-(n-)CNFETs at room
temperature in a high vacuum system (base pressure
$10^{-7}\,$Torr.) Doping is done in-situ using a resistive
potassium source.\cite{dai_apl,vd_apl} We use short K-deposition
steps ($1$-$2\,$min) followed by brief ($30\,$min) anneals at
temperatures between $120$-$170\,^{\circ}$C. Annealing temperature
in this range significantly increases the mobility of K on the
substrate and allows quasi-uniform distribution on the nanotube,
as has been shown in the case of other
fullerenes.\cite{poirier_science91} After each device has been
exposed to some number of doping/annealing cycles (typically a
{\it total} of $10$-$15\,$min of K vapor and $4$-$16\,$hours of
annealing), it arrives at a final state where performance can not
be improved with further processing. Additional benefit of
annealing after deposition is that it prevents gross overcoating
of the nanotube, as well as the area between the source and the
drain with potassium.\cite{bock_prb} The observed improvement
cannot be achieved by doping alone which only shifts the threshold
voltage toward negative gate values ($V_{\rm gs}$) without any
increase in electron current (data not shown).

Next we compare the performance of the as-prepared CNFETs with the
device characteristics resulting following the exposure to
K-doping and annealing. Figure~\ref{Fig1}(a) shows initial and
final transfer characteristics ($I_{\rm d}$ vs.~$V_{\rm gs}$) of
two different devices converted in two different runs. The final
devices show two clearly apparent differences compared to the
as-prepared CNFET: (i) the $V_{\rm th}$ for electron conduction
shifts from about $0.5\,$V in (nearly) intrinsic CNFET to
$-1.3\,$V, and (ii) the on-current for electrons is increased by
$3$-$4$ orders of magnitude compared to the initial electron
current and is $10$-$100$ times higher then initially more
pronounced hole current. This second effect is a direct
consequence of the annealing of the device and constitutes the
main experimental finding of this work.

Beyond the outlined changes, our CNFETs behave as excellent
depletion mode n-type transistors. In particular, the inverse
subthreshold slope, $S=dV_{\rm gs}/d(\log{I_{\rm d}}) \le
150\,$mV/decade approaches the thermal limit.  This is because of
the improved electrostatics due to the thin $5\,$nm oxide which
affords great control over the nanotube bands such that tunneling
through the SB\cite{ja_SBprl,rm_SBprl,sh_SBprl} proceeds with
nearly perfect transmission.\cite{guo}  Throughout the K exposure
and annealing, the device maintains a steep subthreshold slope and
a large on-off ratio $5$-$6$ (orders of magnitude) even at $V_{\rm
ds} \sim 1.0\,$V, which is due to the small diameter and
correspondingly rather large band gap of our
nanotubes.\cite{guo,mr_apl} Another potential advantage of CNFETs
if compared to Si-based devices is the predicted identical
performance of p- and n-transistors due to same values for hole
and electron effective masses. Figure~\ref{Fig1}(b) shows the
output characteristic ($I_{\rm d}$ vs.~$V_{\rm ds}$) of the
improved n-CNFET in the intermediate conversion in
Fig.~\ref{Fig1}(a). Even at this stage currents in excess of
$1\,\mu$A are driven through the device, corresponding to a large
transconductance $g_{\rm m} = dI_{\rm d}/dV_{\rm gs} \sim
1\,\mu$S. Further K exposure and annealing results in improvement
by about a factor of $3$-$5$ in $g_{\rm m}$.\cite{vd_foot} These
values are comparable to the very best small diameter
p-CNFETs,\cite{sw_apl} in line with the expectations based on the
band structure. All these observations together suggest that
device performance can be optimized in {\it both} on- and
off-states without sacrificing the tunability of $V_{\rm th}$.

\begin{figure}
\begin{center}
\epsfig{file=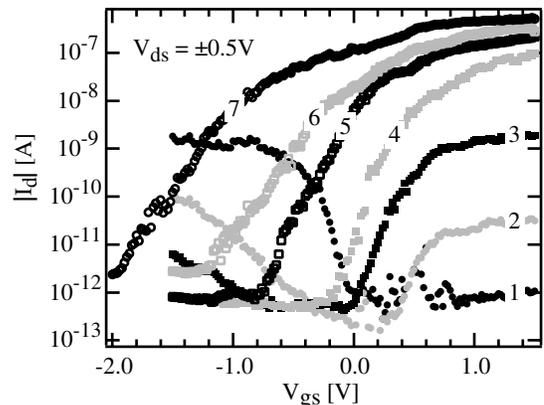} \caption{Stepwise conversion and
improvement of a CNFET. Initial p-type curve (\#1) is followed by
one with ambipolar characteristics (\#2) due to vacuum annealing.
Consecutive doping and annealing steps (\#3-\#7) gradually
transform the device into an n-CNFET by increasing the on-current
and shifting the threshold voltage to more negative values.}
\label{Fig2}
\end{center}
\end{figure}

Given these promising device characteristics, we would like to
understand the mechanism of this improvement by looking at a
stepwise doping and annealing cycle such as shown in
Figure~\ref{Fig2}. The initial p-CNFET (curve \#1) is converted to
an ambipolar device (curve \#2) by an overnight vacuum anneal at
$170\,^{\circ}$C.\cite{vd_apl} Curves \#3-7 show 5 consecutive
short doping/annealing steps all of which exhibit {\it both} a
shift in the threshold voltage and an increase in the on-current
compared to previous characteristic. This implies that potassium
exposure and heating treatment are responsible for both effects.
However, if potassium only transfers charge to the nanotube by
chemically doping the body of the nanotube (device bulk), this
would explain the threshold voltage shift but not the increase in
on-current. For this reason, we have to conclude that the observed
changes are produced by a combination of bulk doping and improved
injection from the source electrode into the CNFET. In the
following we consider each aspect independently.

\begin{figure}
\begin{center}
\epsfig{file=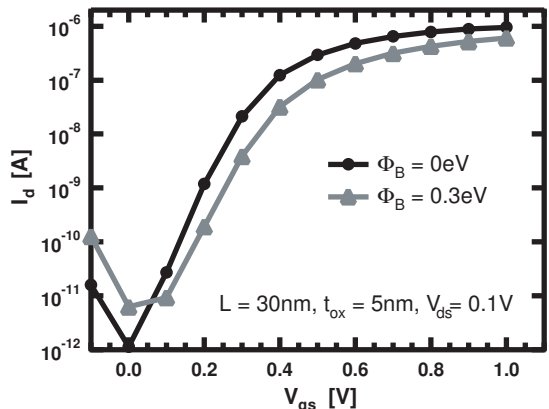} \caption{Transfer characteristics for
different SB heights of a typical small diameter nanotube in our
device geometry modelled by a fully self-consistent, quantum
mechanical simulations (see Ref.~[20]). It is evident that the
on-current is initially (at midgap line-up) much higher than in
the experiment and that it increases by less than an order of
magnitude with lowering of the SB height.} \label{Fig4}
\end{center}
\end{figure}

In our previous work we have focused on the off-state of nanotube
transistors \cite{ja_SBprl,sh_SBprl}. Here the emphasis is on how
the on-current $I^{\rm on}_{\rm d}$ of a CNFET is affected by
potassium exposure and annealing treatment. First, we evaluate the
performance of the initial, undoped device. $|I^{\rm on}_{\rm d}|$
is typically in the nA range [see Figs.~\ref{Fig1}(a)
and~\ref{Fig2}] at drain voltages V$_{\rm ds} = -0.5\,$V. In
comparison with the {\it expected} response from self-consistent
quantum mechanical simulations on the performance of Schottky
barrier (SB) CNFETs\cite{knoch03} this current is at least one
order of magnitude too small - see Fig.~\ref{Fig4}. This statement
is true no matter what Schottky barrier height $\Phi_{\rm B}$ is
assumed for a gate oxide thickness of t$_{\rm ox} = 5\,$nm. Other
groups have also pointed out that $|I^{\rm on}_{\rm d}|$ becomes
progressively less affected by $\Phi_{\rm B}$ the stronger the
gate impact (e.g. the thinner the gate dielectric).\cite{guo} At
the same time the n-type devices after K-deposition show regularly
on-currents that are in the micro-amps range as expected for our
device geometry. Since the increase in on-current between devices
before and after K-doping cannot be explained as a result of a
barrier lowering {\it and} $|I^{\rm on}_{\rm d}|$ is always too
low for the initial CNFETs, we conclude that an additional contact
resistance is present before K-deposition.\cite{ja_sub}

In order to {\it consistently} explain our experimental
observations we have to assume that this extra series resistance
limits the device performance in the {\it on-state} prior to the
potassium treatment and has vanished or substantially decreased
afterwards. This observation is in agreement with the reports that
the metal/nanotube contact resistance for metals having similar
work functions can be substantially different.\cite{dai_nature} It
is believed that this effect is related to the varying ability of
different metals to wet the nanotube surface.\cite{zhang00} Our
results indicate that the intimacy between the contacting metal
and the nanotube changes when the titanium/nanotube interface is
replaced by a potassium/nanotube contact as is the case after
K-doping and annealing. It should be noted that this gate
field-independent contact resistance {\it is NOT} a replacement
for the formerly introduced Schottky barrier (SB) model. SBs have
to be assumed to explain in particular the transistor {\it
off-state} characteristics.\cite{ja_SBprl,sh_SBprl} Instead, our
findings here suggest that {\it in addition} a field independent
parasitic resistance has to be taken into account using certain
metal contacts to explain the transistor {\it on-state}.

Next we discuss the correspondence between the shift in the Fermi
level of the CN due to K-deposition and the movement of the
$V_{\rm th}$ compared to an undoped device.  Since the gate
dielectric is quite thin ($t_{\rm ox} = 5\,$nm) we can to first
order estimate the change in the Fermi level, at any value of SB
height, to be the same as the change in the threshold voltage:
$\Delta E_{\rm F} = q \Delta V_{\rm th}$. In Fig.~\ref{Fig2}, the
Fermi level shifts by about $1\,$eV from the intrinsic (\#2) to
the fully doped curve (\#7), implying that it has moved $\sim\,
500\,$meV into the conduction band.  The inferred doping level is
quite a bit higher than in bulk Si where the highest doping is to
$\sim\, 100\,$meV above the conduction band edge (and $\sim\,
30\,$meV below the valence band edge) due in part to ineffective
screening in one-dimension (1D).

Finally, we try to use the estimated Fermi level position to
extract the actual linear charge density transferred from K vapor
by computing the 1D density of states for a parabolic band. By
accounting for both bands and spin degrees of freedom in a CN and
omitting Fermi distribution function to simplify the integral we
obtain $n_{\rm 1d} = (\sqrt{8m^*(E_{\rm F} - E_{\rm c})})/(\pi
\hbar)$, where $m^*$ is the effective mass, and $(E_{\rm F} -
E_{\rm c})$ is the position of the Fermi level compared to the
band edge ($\le\, 500\,$meV in our experiment).  Substituting
numerical values $n_{\rm 1d} \sim 3.2\,{\rm
nm}^{-1}\sqrt{(m^*/m_0)(E_{\rm F} - E_{\rm c})[eV]}$ and given
$(m^*/m_0) \sim 0.06$ for CNs of this diameter, we obtain $n_{\rm
1d} \sim 0.6\,{\rm nm}^{-1}$. This is indeed a high linear doping
density corresponding to a bulk density $n_{\rm 3D} \sim
10^{20}\,$cm$^{-3}$. As there are approximately 180 carbon
atoms/nm of length in small diameter CNs, this linear density
converts to stochiometric formula of KC$_{300}$ assuming that each
K atom donates one full electron.\cite{jo_prb02} The doping level
is similar to previously computed densities in other doped
CNFETs,\cite{bock_prb} but far below the charge transfer inferred
in optical experiments.\cite{opto1}

In summary, we have developed a technique which reproducibly
yields high performance (on-current 5-$6\,\mu$A) n-CNFETs by
utilizing K-doping/annealing to: (i) increase the charge density
in the channel (up to $0.6\, e^-/$nm), (ii) reduce external
resistances (and likely SB height) in the device. Doping/annealing
cycles give the ability to tune the threshold voltage over a wide,
technologically relevant range ($-1.3\,$V$\le V_{\rm th} \le
0.5\,$V). Our CNFETs benefit from using large band gap/small
diameter nanotubes and vertical scaling to ensure excellent
off-state, switching and subthreshold swing ($S \le
150\,$mV/decade). Together, these features make the devices
reported here on par with {\it any} of the best p-CNFETs
fabricated to date,\cite{sw_apl,dai_nmat,dai_nature} and most
appropriate for integration into nanotube-based complimentary
logic gates.

%
%

We wish to thank J. Buccignano for e-beam exposures, and B. Ek for
expert technical assistance.

\bigskip

\end{multicols}

\begin{references}
\bibitem{vd_foot} It is worth noting that this magnitude of on-current in K-doped n-CNFETs
has been previously reported on devices with thicker gate oxide
and unoptimized subthreshold swing.\cite{vd_nl}
\end{references}
\end{document}